\documentclass{iucrjournals}

\usepackage{siunitx}

\title{A low-background setup for \emph{in-situ} X-ray total scattering combined with fast scanning calorimetry}

\author[a]{Peihao Sun\IUCrCemaillink{peihao.sun@unipd.it}\IUCrOrcidlink{0000-0002-3608-1557}}%
\author[a]{Jacopo Baglioni\IUCrOrcidlink{0000-0003-0322-0941}}%
\author[a]{Beatrice Baraldi}%
\author[a]{Weilong Chen}%
\author[a]{Daniele Lideo\IUCrOrcidlink{0009-0004-8397-5264}}%
\author[a]{Lara Piemontese}%
\author[a]{Francesco Dallari\IUCrOrcidlink{0000-0003-0171-8084}}%
\author[b]{Marco Di Michiel}%
\author[a]{Giulio Monaco\IUCrOrcidlink{0000-0003-2497-6422}}%

\affil[a]{Dipartimento di Fisica e Astronomia ``Galileo Galilei'', Universit\`a degli Studi di Padova, Via F.\ Marzolo 8, 35131 Padova, Italy}
\affil[b]{ESRF, The European Synchrotron, 71 Avenue des Martyrs, CS40220, Grenoble, 38043 Cedex 9, France}

\begin{document} 
\maketitle 

\begin{synopsis}
A setup enabling \emph{in-situ} X-ray total scattering combined with fast scanning calorimetry is demonstrated at beamline ID15A of the ESRF, allowing for the retrieval of detailed structural evolution with changes in thermodynamic conditions.
\end{synopsis}

\begin{abstract}
We demonstrate a setup combining fast scanning calorimetry with X-ray total scattering at a synchrotron beamline, allowing for \emph{in-situ} characterizations of the nano-scale structure of samples during and after temperature scans. The setup features a portable vacuum chamber giving high signal-to-background ratio even on amorphous samples, which enables the observation of detailed structural changes between different sample states. We show three use cases, including one which leverages the high cooling rate of 10$^4$\,K/s achievable by this setup. Our demonstration opens the door to various applications in materials science where it is important to understand the interplay between structure and thermodynamics.
\end{abstract}

\keywords{fast scanning calorimetry; X-ray total scattering; structure of glasses; in-situ structural measurements}

\section{Introduction}
Obtaining atomic-scale structural information of materials under different thermodynamic conditions is crucial to the understanding of material properties and the physical mechanism governing their behaviors. This is particularly important for systems like glasses whose properties can depend sensitively on their thermal history (for example, the cooling rate at which they are quenched down from the liquid state)~\cite{Debenedetti2001}. Therefore, the ability to perform \emph{in-situ} structural measurements on samples during and after controlled thermal programs is highly valuable for studies in the field of physics and materials science.

The temperature control of samples for \emph{in-situ} measurements is typically realized through the use of conventional furnaces based on radiative or conductive heating. These devices often suffer from serious inaccuracies in the determination of sample temperature or from very limited temperature ramping rates due to large thermal masses~\cite{Doran2017CompactDiffraction}. More recent works have improved on the design, resulting in furnaces which allow for \emph{in-situ} structural characterizations with higher temperature precision and higher ramping rates~\cite{Doran2017CompactDiffraction,Chakraborty2015DevelopmentSpectromicroscopy,Marshall2023AHeater}. Still, the heating and cooling rates are limited to \SI{\sim 100}{K/s} and more often to \SI{\sim 5}{K/s}, and in some designs, the need for a gaseous environment for efficient thermal equilibration leads to significant additional background in structural measurement techniques including X-ray diffraction.

Nanocalorimetry, including fast scanning calorimetry (FSC)~\cite{Mathot2023,Yi2019Nanocalorimetry:Smaller}, offers a promising alternative for \emph{in-situ} structural and thermodynamic studies of various systems. Due to the small sample mass on the \si{\micro\gram} or even \si{ng} level, nanocalorimeters can achieve heating and cooling rates in excess of \SI{e6}{K/s}~\cite{Mathot2023,Yi2019Nanocalorimetry:Smaller}. In addition, microelectromechanical systems
(MEMS)-based calorimeters~\cite{Mathot2023,Mathot2011,Zhuravlev2010FastDevice}, where the sample is mounted on a sensor that can be placed external to the bulk of the main device, provides a versatile solution that can be combined with X-ray scattering instruments at synchrotron radiation facilities~\cite{Rosenthal2014High-resolutionCalorimetry,Narayanan2017RecentMatter}. For example, in a previous work, we have demonstrated the combination of FSC with X-ray photon correlation spectroscopy and showcased, among other applications, the use of FSC as a fast-responding furnace for dynamic scattering measurements~\cite{Martinelli2024}.

The small sample size for FSC, usually no more than a few tens of \si{\micro\meter} and sometimes even smaller, could in principle present a challenge for nano-scale structural characterizations. However, thanks to recent advancements in synchrotron radiation facilities, it is now possible to have high-flux, high-photon-energy X-ray beams with a small focus, thereby enabling high-quality X-ray total scattering measurements to retrieve the atomic pair distribution function (PDF), providing crucial insight into the atomic-scale structural properties of various materials~\cite{Billinge2019TheEvents}. For example, after the upgrade to the Extremely Brilliant Source (EBS), at beamline ID15A of ESRF - The European Synchrotron, it is possible to have \SI{69}{keV} X-ray beams focused down to \SI{0.3}{\micro\meter}\,$\times$\,\SI{0.3}{\micro\meter} with a flux well exceeding \num{e12} photons per second~\cite{Vaughan2020ID15AScattering}. This makes it possible to perform high-speed \emph{operando} X-ray total scattering experiments on small samples, including those for chip-based FSC.

Therefore, in this work, we demonstrate the combination of X-ray total scattering and FSC for \emph{in-situ} measurements. In particular, we present a setup including a custom-designed low-background sample chamber that enables the collection of X-ray diffraction data with a high signal-to-background ratio. We show that, with this setup, it is possible to obtain high-quality PDFs revealing detailed differences in the sample structure during and in between FSC scans. The rest of this article is organized as follows: Section 2 describes our experimental setup with a quantitative analysis of the background level. Sections 3 to 5 give three example applications of this setup: measurements of detailed structural changes during temperature ramps, quantification of X-ray beam-induced heating effects, and characterizations of atomic structure with different cooling rates spanning 4 orders of magnitude.

\section{A low-background sample chamber for \emph{in-situ} measurements}
\begin{figure}
    \centering
    \includegraphics[width=0.95\textwidth]{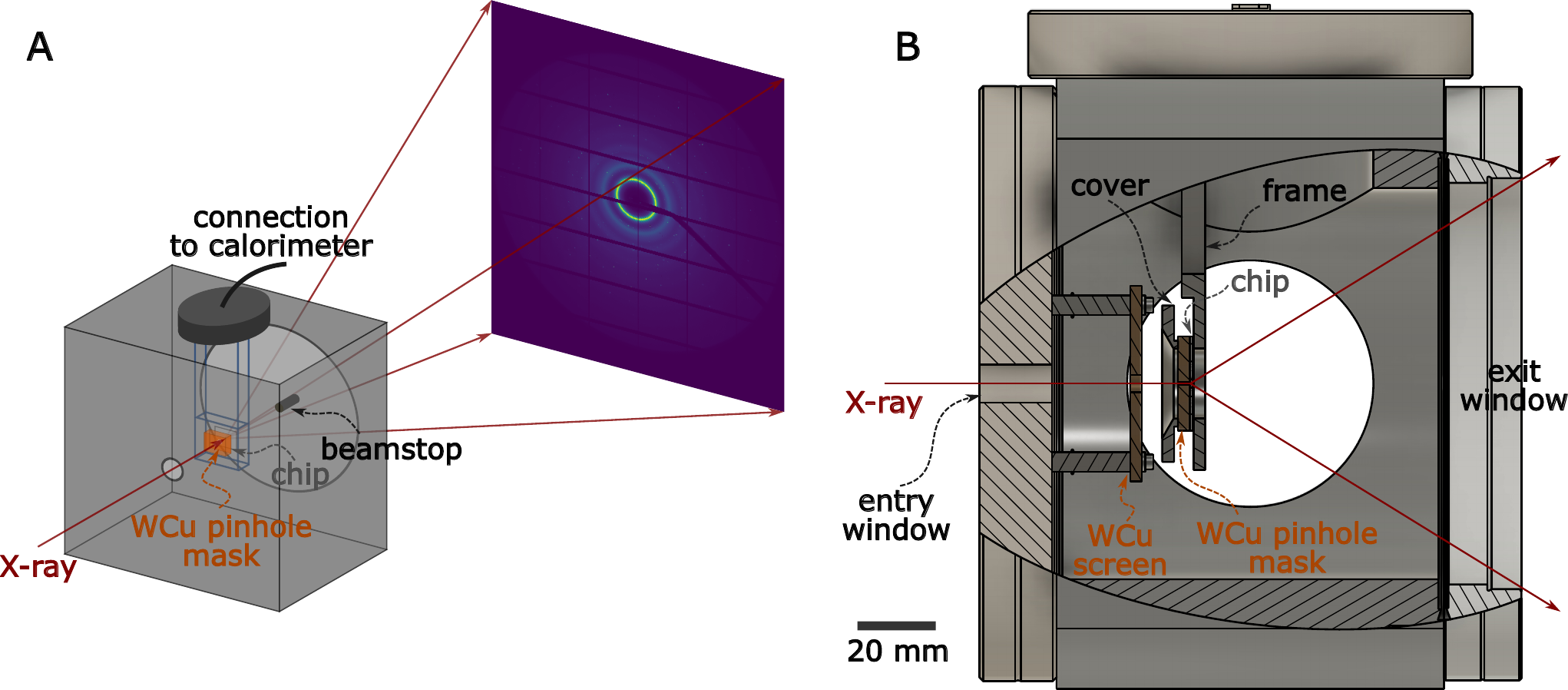}
    \caption{(A) Schematic diagram of the low-background sample chamber (not to scale). The sample is mounted on a FSC chip inside the chamber. A mask made of tungsten alloy is placed before the chip and acts as a pinhole. The chamber features a large exit window of kapton foil allowing measurements up to high $Q$. (B) Section view of the chamber showing various components along the X-ray beam path.}
    \label{fig:setup}
\end{figure}

Figure~\ref{fig:setup} shows a schematic diagram of the setup, including several details of the sample chamber. This setup is realized at beamline ID15A of the ESRF, and a commercial Mettler-Toledo Flash DSC 2+ calorimeter is used. X-ray scattering patterns are collected by a hybrid photon-counting pixel detector Pilatus3 X CdTe~\cite{Vaughan2020ID15AScattering}. Although the sample chamber is designed to be compatible with ID15A, it is made to be stand-alone and can be installed at other beamlines for X-ray diffraction measurements. The chamber is vacuum-compatible, but it can also be filled with gas if necessary (albeit with an increased background level for X-ray diffraction measurements). Because we have demonstrated before that FSC scans can be performed with the sample in vacuum achieving high heating and cooling rates~\cite{Martinelli2024}, in this work the FSC scans and X-ray scattering measurements are done with the chamber under vacuum.

The design of the chamber is similar to that in our previous work~\cite{Martinelli2024}, although with several important modifications to make it a versatile sample environment for implementation at high-energy X-ray scattering beamlines. Firstly, it features a large exit window to allow measurements up to high values of the momentum transfer, $Q=(4\pi/\lambda)\sin\theta$, where $\lambda$ is the X-ray wavelength and $\theta$ is half of the scattering angle. For the demonstrations in this work, the X-ray photon energy is fixed at \SI{67.808}{keV} (with \SI{0.25}{keV} bandwidth), and the valid $Q$-range is from \SIrange{0.5}{19}{\per\angstrom}, with the lower limit given by the \SI{2}{mm}-diameter beamstop placed close to the exit window. The beamstop is mounted external to the chamber and is aligned independently. Secondly, for the chamber to be stand-alone (i.e., not directly connected to the beamline's X-ray pipeline), it features a small window (\SI{10}{mm} in diameter) for incoming X-rays, as indicated in Figure~\ref{fig:setup}. In order to reduce the background that is inevitably created by this window and the air path before it, we have manufactured a mask out of a \SI{3}{mm} thick tungsten alloy sheet (composition W$_{80}$Cu$_{20}$ by wt.\%), with a hole of diameter \SI{\sim 400}{\micro\meter} acting as a pinhole, referred to as ``WCu pinhole mask'' hereafter. This takes advantage of the fact that the calorimeter chips have small active areas (no more than a few hundred \si{\micro\meter} laterally) at fixed positions where the sample can be mounted. Another \SI{50}{mm}\,$\times$\,\SI{50}{mm} square mask with a \SI{5}{mm} diameter hole (``WCu screen'' in Figure~\ref{fig:setup}B), made of the same tungsten alloy, is placed between the entry window and the sample to further reduce the background.

To quantify the background level, we perform X-ray scattering measurements on and off a metallic glass sample of composition Pd$_{42.5}$Cu$_{30}$Ni$_{7.5}$P$_{20}$ (at.\%) and thickness on the order of \SI{\sim 20}{\micro\meter}. The sample is mounted on a Mettler-Toledo UFH1 chip, whose active area consists of gold and SiN$_\text{x}$ layers~\cite{Mathot2023} and has a diameter on the order of \SI{90}{\micro\meter}. Further details of the chips available with Mettler-Toledo Flash DSC machines can be found in previous reports~\cite{Mathot2023,Mathot2011}. In our setup, the chip is fixed together with the WCu pinhole mask onto the sample frame by an aluminum cover (see Figure~\ref{fig:setup}B). Both the cover and the frame have clearances with large diameters ($>\SI{15}{mm}$) at the center for X-rays to pass through. For consistency, the same sample is used here and in the rest of this work, and the X-ray spot size at the sample position is about \SI{1}{\micro\meter}\,$\times$\,\SI{2}{\micro\meter}. Azimuthal integration of the diffraction patterns is performed using the \texttt{pyFAI} library~\cite{Kieffer2025ApplicationCrystallography}, which takes into account the X-ray polarization factor and the solid angle of each detector pixel. The experimental geometry is calibrated using a Cr$_{2}$O$_{3}$ powder fixed onto the chip, and the sample-detector distance is determined to be \SI{199.6}{mm}. We note that, while the detector is mounted on a motor and can be moved if needed, the same detector position is used for all measurements presented below.

At ID15A, the incoming X-ray flux at the sample position can be attenuated by inserting a SiO$_2$ wedge in the beam path upstream from the sample, in experimental hutch EH2~\cite{Vaughan2020ID15AScattering}. The wedge is mounted on a motorized linear stage and can be moved to change its thickness in the beam path, thereby achieving the desired attenuation. The correspondence between the motor position and the X-ray flux onto the sample is calibrated using a beam intensity monitor (PIN diode). The attenuation does not change the beam shape at the sample position since the homogeneous chemical composition and optically polished surfaces of the wedge do not alter the shape of the X-ray beam wave front.

\begin{figure}
    \centering
    \includegraphics{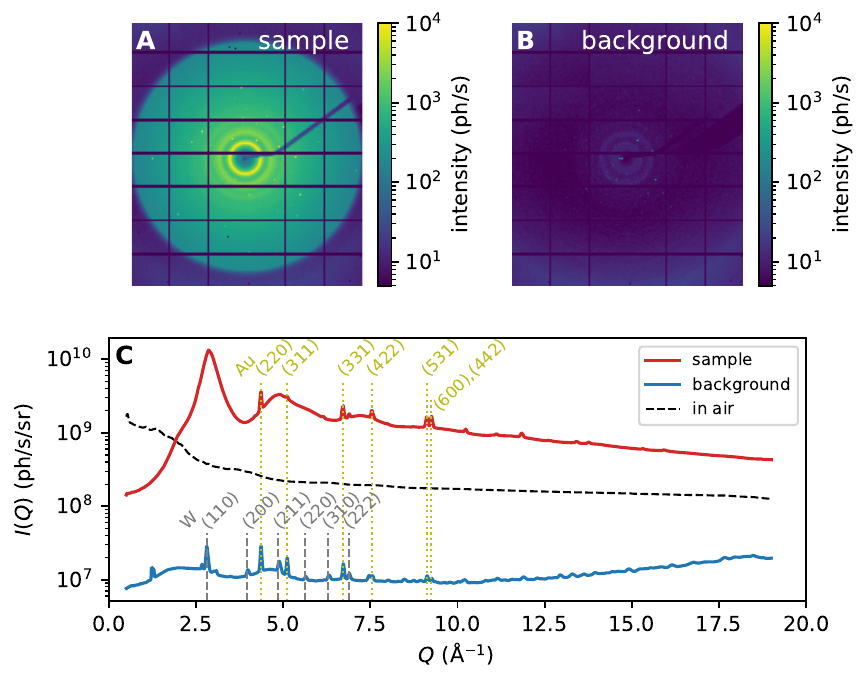}
    \caption{Background characterization. (A) Example diffraction pattern from a Pd$_{42.5}$Cu$_{30}$Ni$_{7.5}$P$_{20}$ metallic glass sample, roughly \SI{20}{\micro\meter} thick, mounted on a Mettler-Toledo UFH1 chip. (B) Background diffraction pattern taken at another position on the chip away from the sample, plotted with the same color map. (C) Azimuthally integrated scattering intensity profiles for the sample (red) and background (blue), in units of photons per second per unit solid angle. The dashed black line shows the intensity profile with the chamber in air and without the sample assembly including the WCu pinhole mask (see text). Dashed and dotted vertical lines indicate the positions of Bragg reflections of tungsten (W) and gold (Au), respectively, that are visible in the intensity profiles.}
    \label{fig:background}
\end{figure}

Figure~\ref{fig:background}A shows an example image of the diffraction pattern on the sample, where the color map is set on a log scale in order to show different features in the pattern. The incoming X-ray flux is about \num{4.5e11} photons per second in this measurement. Scattering signal up to at least the third diffraction peak can be identified in the image, while the large circular contour corresponds to the edge of the exit window. In Figure~\ref{fig:background}B, we show the background scattering taken at a position away from the sample, with the same incoming X-ray flux and using the same color map. It is evident that the intensity here is orders of magnitude lower than the signal from the sample. To show this more clearly, we plot the integrated intensity profiles in Figure~\ref{fig:background}C. It can be seen that the background is negligible compared with the signal from the sample, even in the low-$Q$ region where the structure factor of the sample is expected to be very small. The profile subtracting the background from the signal is not shown here because it visibly overlaps with the total signal shown in the plot (more information is provided in Appendix~\ref{sec:app:data_processing}). The two broad peaks in the background centered around \SI{2}{\per\angstrom} and \SI{4.6}{\per\angstrom}, as can also be seen in Figure~\ref{fig:background}B, are consistent with the structure factor of amorphous silicon nitride~\cite{Aiyama1979AnNitride} and arise from the silicon nitride substrate of the FSC chip. In comparison, the black dashed line in Figure~\ref{fig:background}C indicates the background level after the chamber was opened to ambient pressure and the sample assembly, including the WCu pinhole mask, was removed. Note that for this measurement, it was necessary to move the chamber upstream by \SI{11}{mm} to prevent the kapton exit window from touching the beamstop when it expands upon venting the chamber. It is clear that the background level becomes much higher, particularly in the low-$Q$ region. This kind of background can be difficult to account for accurately, as it contains contributions from different sources (for example, air scattering before and after the sample) that will be attenuated differently after the insertion of the sample.

It can be seen in Figure~\ref{fig:background} that some sharp Bragg peaks appear in the diffraction images. As shown in Figure~\ref{fig:background}C, these Bragg peaks can be attributed to tungsten and gold. The tungsten peaks arise from the tungsten alloy mask and are negligible compared to the scattering signal from the sample, as shown in Appendix~\ref{sec:app:data_processing}. The gold peaks arise from the active area and from the wiring of the UFH1 chip, so they are not intrinsic to the chamber design but are specific to the chip we used. These peaks are stronger at the sample position where the X-ray beam directly hits the active area, and their intensity varies with position due to the changing texture and thickness of the gold layer. During sample alignment, we typically adjust the beam position on the sample to minimize the intensity of these peaks. Then, in data processing, we identify the remaining peaks and mask the surrounding pixels; the procedure is described in Appendix~\ref{sec:app:data_processing}.

\section{\emph{In-situ} X-ray total scattering during temperature ramps\label{sec:T_ramp}}
\begin{figure}
    \centering
    \includegraphics{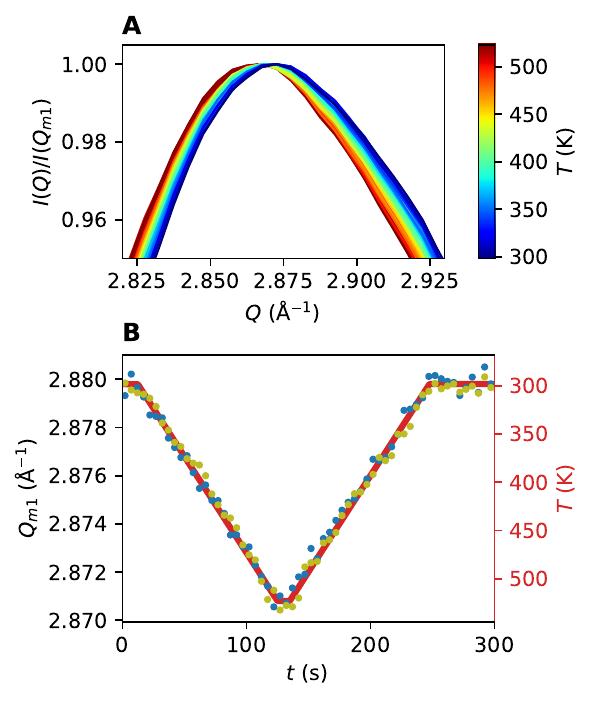}
    \caption{Demonstration of \emph{in-situ} X-ray diffraction during an FSC scan. (A) The intensity profile $I(Q)$ normalized by its maximum value, focusing on the vicinity of the first sharp diffraction peak to show its temperature dependence during the heating and cooling ramps of a scan. From dark blue to dark red, the curves show increasing temperatures, as indicated by the color bar. (B) Blue and yellow dots show the center position of the first sharp diffraction peaks, $Q_{m1}$, as a function of time during two different scans. The temperature program is shown as the red line (right $y$-axis).}
    \label{fig:T_ramp_Qm1}
\end{figure}

As a first demonstration, we show that the setup can be used as a fast-responding furnace that allows for quick and accurate measurements of the atomic structure at various temperatures. To do this, we set up the following temperature scan program on the calorimeter: a temperature ramp up from \SI{298}{K} to \SI{523}{K} at \SI{2}{K/s}, a \SI{10}{s} isotherm at \SI{523}{K}, and a symmetric ramp down to \SI{298}{K} at \SI{2}{K/s}. The scan is padded with isotherms at \SI{298}{K} at the beginning and the end of the temperature ramps to allow for possible mismatches in launching the X-ray measurements. The highest temperature is kept well below the glass transition temperature $T_g \approx \SI{580}{K}$~\cite{Georgarakis2011VariationsVitrification} in order to avoid aging effects above $T_g$ that can obscure the reproducibility between scans. A visual representation of the temperature program is shown as the red line in Figure~\ref{fig:T_ramp_Qm1}B (right $y$-axis). Then, we collect X-ray diffraction patterns during the scan. Here, the incoming X-ray flux is about \num{4.4e10} photons per second, which does not cause a significant temperature rise as will be demonstrated in the next section. The frame rate is \SI{1}{Hz}, and the exposure time is \SI{0.99}{s} per frame. We repeat the scan twice to check the reproducibility of the results.

A sensitive and direct observable that reflects temperature change in this sample is the position of the first sharp diffraction peak, $Q_{m1}$. This can be seen in Figure~\ref{fig:T_ramp_Qm1}A, which show the azimuthally integrated intensity profile $I(Q)$ normalized by its maximum value $I(Q_{m1})$ during the heating and cooling ramps of a scan. Each curve represents an average over 5 images.
When extracting $Q_{m1}$ from these data, since the curves do not appear to have an exact functional shape (e.g., Gaussian), and the differences between them are small, the absolute values of $Q_{m1}$ may depend somewhat on the model used to extract them. Thus, to avoid ambiguity, we define $Q_{m1}$ as the first moment (center of mass) of $I(Q)$ in the range where it is above 95\% of its maximum value, as shown in Figure~\ref{fig:T_ramp_Qm1}A, which we find to be a definition that is robust and easy to implement.
We then plot $Q_{m1}$ as a function of time during the scans, shown as dots in Figure~\ref{fig:T_ramp_Qm1}B. The two colors represent the two different scans, and the good agreement between them demonstrates a high level of reproducibility. Furthermore, the shift in $Q_{m1}$ overlaps very well with the temperature profile, showing the possibility to use the calorimeter as a fast-responding and accurate furnace.

\begin{figure}
    \centering
    \includegraphics{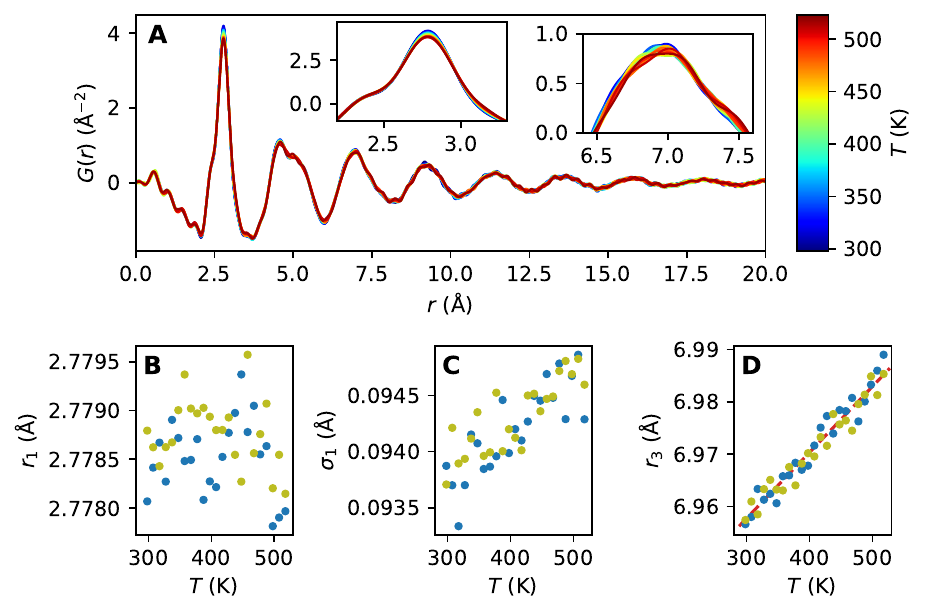}
    \caption{PDF analysis as a function of temperature during the heating ramp. (A) The full $G(r)$ profile. The insets show magnified views of the 1$^\text{st}$ and 3$^\text{rd}$ peaks. The lower panels show, as a function of temperature in two different scans: (B) the mean position of the 1$^\text{st}$ peak, $r_1$; (C) the standard deviation of the 1$^\text{st}$ peak, $\sigma_1$; (D) the mean position of the 3$^\text{rd}$ peak, $r_3$. The red dashed line in (D) shows a linear fit of the data.}
    \label{fig:T_ramp_PDF}
\end{figure}

One of the main advantages of our setup is its capability of covering a wide $Q$-range with a low background, which enables the reconstruction of the atomic PDF of the sample \emph{in situ} during temperature scans. Therefore, we perform PDF analysis on the diffraction data using the software \texttt{PDFgetX3}~\cite{Juhas2013}, and the results are shown in Figure~\ref{fig:T_ramp_PDF}A (for clarity, only data from the temperature ramp up of one of the scans is shown in this panel). Even though the changes are small as expected for this temperature range~\cite{Georgarakis2011VariationsVitrification}, they are nonetheless clearly visible. For example, as shown in the insets in Figure~\ref{fig:T_ramp_PDF}A, with increasing temperature, the 1$^\text{st}$ peak (centered around \SI{2.8}{\angstrom}) lowers in amplitude and becomes broader without apparently shifting its center position, while the 3$^\text{rd}$ peak (centered around \SI{7.0}{\angstrom}) shifts to higher $r$. Higher order peaks behave similarly as the third peak, so here we focus on the 3$^\text{rd}$ peak due to its larger amplitude giving higher signal-to-noise ratio.

To quantify the observations above, we extract and plot the following quantities from the $G(r)$ data: the first moment of the main 1$^\text{st}$ peak, $r_1$ (Figure~\ref{fig:T_ramp_PDF}B), the square root of the second moment of the 1$^\text{st}$ peak (i.e., its standard deviation), $\sigma_1$ (Figure~\ref{fig:T_ramp_PDF}C), and the first moment of the 3$^\text{rd}$ peak, $r_3$ (Figure~\ref{fig:T_ramp_PDF}D). We choose the following range to calculate these quantities: the range for the 1$^\text{st}$ peak of $G(r)$ is determined by the two near-isosbsetic points at around \SI{2.60}{\angstrom} and \SI{2.95}{\angstrom}, while the range for the 3$^\text{rd}$ peak is determined by the roots of $G(r)$ around \SI{6.5}{\angstrom} and \SI{7.5}{\angstrom} (varying slightly with temperature). Note that the results depend little on the exact range chosen. We can see that Figure~\ref{fig:T_ramp_PDF}B-D quantitatively confirm the observations above. In particular, $r_3$ appears to expand linearly with temperature, as shown by the fit (red dashed line) in Figure~\ref{fig:T_ramp_PDF}D. The slope suggests a linear expansion coefficient of \SI{1.8e-5}{\per\kelvin}, or a volumetric expansion coefficient of \SI{5.3e-5}{\per\kelvin}, consistent with reported values in the literature~\cite{Lu2002InvestigationState,Georgarakis2011VariationsVitrification}. We note that similar values can be obtained from the slope of $Q_{m1}$ shown in Figure~\ref{fig:T_ramp_Qm1} above. We also note in passing that the much smaller changes (if any) in $r_1$ and the increase in $\sigma_1$ suggest that the overall thermal expansion is not related to changes in the average nearest-neighbor distance, but to an increase in the width of its distribution. In other words, in this temperature range, the system appears to expand in the long range accompanied by an increase in local disorder.

\section{Quantifying X-ray beam-induced heating}
\begin{figure}
    \centering
    \includegraphics{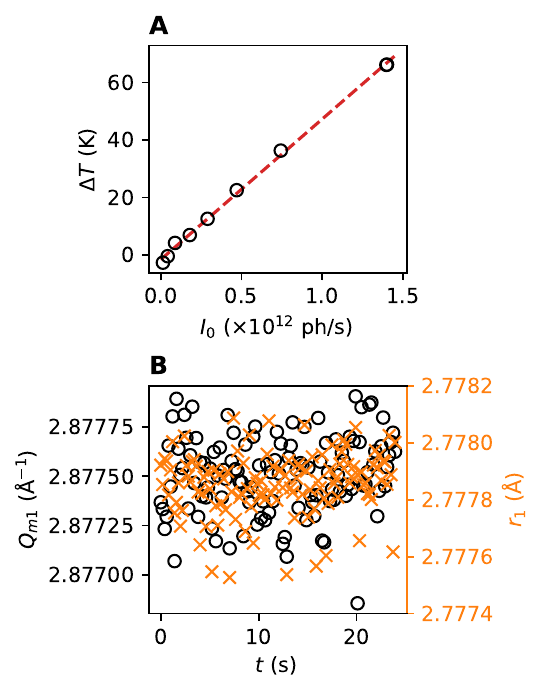}
    \caption{(A) Effective temperature rise due to X-ray beam. Data points (black circles) are calculated by extracting the difference of $Q_{m1}$ at each incoming flux with respect to $Q_{m1}$ at lowest flux (0.3\% transmission, about \num{\sim 4.2e9} photons per second), and converted to $\Delta T$ using the linear relation obtained from Figure~\ref{fig:T_ramp_Qm1}. The red dashed line shows a linear fit. (B) $Q_{m1}$ position (black circles, left $y$-axis) and $r_1$ position (orange crosses, right $y$-axis) as a function of time under full beam. No significant changes are observed.}
    \label{fig:fluence}
\end{figure}

As an application of this setup, we characterize the effective heating due to the X-ray beam, which is highly relevant for measurements on temperature-sensitive samples. To do this, we use the attenuator available at the beamline to tune the incoming flux, $I_0$, from full beam (\num{\sim 1.4e12} photons per second) to 0.3\% transmission (\num{\sim 4.2e9} photons per second). During this experiment, the synchrotron was operating under the uniform filling mode. Then, we take X-ray diffraction images at each flux value, from which we can extract the position of the first sharp diffraction peak, $Q_{m1}$. Since we have shown in the previous section that $Q_{m1}$ shifts linearly with temperature at least up to \SI{523}{K}, we may infer the effective temperature rise from the $Q_{m1}$ position at each $I_0$ value. Specifically, we obtain the difference between the measured $Q_{m1}$ at each $I_0$ and that measured at the lowest flux (0.3\% transmission), and calculate the corresponding temperature difference $\Delta T$ using the measured slope from Figure~\ref{fig:T_ramp_Qm1}, $\Delta Q_{m1}/\Delta T = \SI{-4.0e-5}{\angstrom^{-1}/\kelvin}$. The results are plotted in Figure~\ref{fig:fluence}A. We can see that $\Delta T$ scales linearly with the incoming flux, $I_0$, as expected. Importantly, the effective temperature rise with full beam appears to be more than \SI{60}{K}, which may lead to rather serious consequences if not considered carefully (for example, it may cause effective shifts in the apparent glass transition temperature of the sample). This significant temperature rise is likely an effect of the relatively high absorption by the sample ($Z=46$ for Pd) and the small beam size (even though the effective area of energy deposition is larger due to the diffusion of secondary electrons). Nonetheless, the values obtained here may offer as a guide for measurements under other conditions.

One might be concerned about permanent beam-induced structural modifications which are observed in several network glasses~\cite{Ruta2017,Baglioni2024UniquenessYielding}. Here, we show that if any effect exists, it is negligible in this metallic glass sample. This can already be seen in Figure~\ref{fig:T_ramp_Qm1}, where the $Q_{m1}$ returns to its original value after the temperature ramp down and is highly reproducible in a subsequent scan (no other calorimetric scans were performed in between). To further demonstrate the lack of significant beam-induced structural changes, we take diffraction patterns while irradiating the sample with full beam. In Figure~\ref{fig:fluence}B, we plot $Q_{m1}$ (left $y$-axis) as well as $r_1$ (right $y$-axis, extracted from the PDF) as a function of time. Both quantities appears to remain constant within the noise level (note the small ranges of $y$-axes), and no time-dependent trend can be seen. Therefore, we observe no permanent beam-induced effects beyond noise in this sample with more than \SI{20}{s} of irradiation by the full beam, despite the apparent \SI{\sim 70}{K} temperature rise due to X-ray irradiation.

We note that the flux used both in the previous section and in the next section is about \num{4.4e10} photons per second (about 3\% transmission). The results suggest that $\Delta T$ is negligible (less than \SI{3}{K}) under this flux.

\section{Cooling-rate dependence of sample structure}
In this section, we demonstrate an application leveraging the ability of the flash calorimeter to prepare samples at controlled cooling rates spanning several orders of magnitude. To do this, we launch temperature programs in which the sample is first heated up to \SI{643}{K} at a rate of \SI{1000}{K/s}, then kept at \SI{643}{K} with an isotherm of \SI{10}{s}, and finally cooled down at a given cooling rate to \SI{298}{K}. The cooling rates used here range from \SI{1}{K/s} to \SI{e4}{K/s}. We note that, because the isotherm temperature \SI{643}{K} is much higher than the glass transition temperature, $T_g \approx \SI{580}{K}$~\cite{Georgarakis2011VariationsVitrification}, the sample has reached equilibrium at \SI{643}{K} before being quenched. Then, we make X-ray total scattering measurements on the sample thus prepared, using an incoming X-ray flux of about \num{4.4e10} photons per second. We note in passing that the sequence of the cooling rates used are in random order, so as to exclude systematic shifts in the structure with time during the measurements.

\begin{figure}
    \centering
    \includegraphics{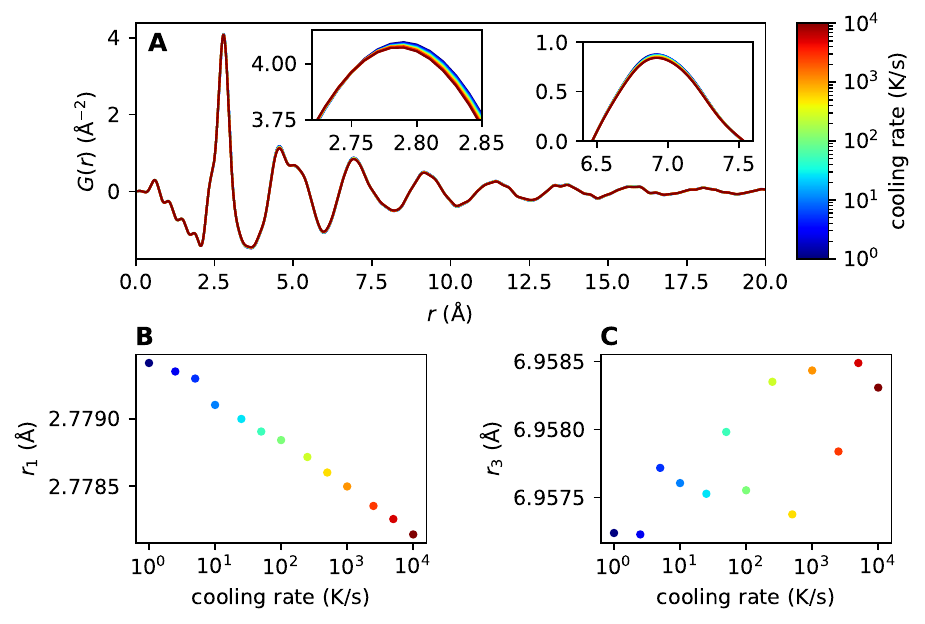}
    \caption{Structural dependence on the cooling rate. (A) PDF of the sample quenched at different rates after equilibration at \SI{634}{K} for \SI{10}{s}. From dark blue to dark red, the curves show increasing cooling rates, as indicated by the color bar. The insets show magnified views of the 1$^\text{st}$ and 3$^\text{rd}$ peaks. (B) The mean position of the 1$^\text{st}$ peak, $r_1$, and (C) the mean position of the 3$^\text{rd}$ peak, $r_3$, as a function of the cooling rate.}
    \label{fig:cooling_rate}
\end{figure}

The PDF obtained from our measurements are presented in Figure~\ref{fig:cooling_rate}, where the curves from dark blue to dark red represent increasing cooling rates. The changes are small but nonetheless visible, as shown in the magnified views in the insets: with higher cooling rates, the 1$^\text{st}$ peak appears to shift inward, while the 3$^\text{rd}$ peak appears to lower in amplitude without shifting significantly in its center position. In Figure~\ref{fig:cooling_rate}B-C, we further quantify these observations by extracting the first moments of the 1$^\text{st}$ peak, $r_1$, and that of the 3$^\text{rd}$ peak, $r_3$, in the same way as detailed in Section~\ref{sec:T_ramp}. Notably, we observe that $r_1$ appears to shift logarithmically with the cooling rate. Since the same trend appears to continue to the highest cooling rate, \SI{e4}{K/s}, it suggests that the cooling rate achievable under vacuum with this sample is at least \SI{e4}{K/s}.

It is particularly noteworthy that the structural changes observed here are rather different from those measured during the temperature ramp as shown in Figure~\ref{fig:T_ramp_PDF}. For example, whereas $r_1$ does not seem to change significantly with temperature up to \SI{523}{K}, here it appears to contract with higher cooling rates, while higher order shells do not seem to change in their positions. We note that this curious contraction of the 1$^\text{st}$ coordination shell is observed also when the sample is heated above $T_g$ in a previous work~\cite{Georgarakis2011VariationsVitrification}. A more detailed understanding of the mechanism behind these structural changes with cooling rate is beyond the scope of this work and can be the subject of future studies.

\section{Conclusions}
We have presented above a setup that combines FSC with X-ray total scattering which enables \emph{in-situ} structural characterizations on the atomic scale during and after FSC scans. Using a low-background vacuum-compatible sample chamber, we are able to collect high-quality diffraction data from a Pd-based metallic glass sample and observe detailed structural changes between different thermodynamic states, including during temperature ramps and after quenching from the supercooled liquid state at different cooling rates ranging from \SI{1}{K/s} to \SI{e4}{K/s}. The results indicate interesting structural behaviors in this sample which can become the basis for future studies, and the high level of structural details revealed demonstrates the potential of the setup to be applied to a wide range of materials studies.

\appendix
\section{Data processing\label{sec:app:data_processing}}
As mentioned above in the main text, the Bragg peaks arising from the gold layer in the active area of the chip may have an influence on the PDF analysis, but it is possible to identify and mask them. Here we describe our algorithm for peak masking. Firstly, we apply a 2D Gaussian filter on the image with $\sigma = 3\;\text{pixels}$ in each dimension. Then, we identify the pixels in which the ratio between the original and the filtered image, $I_\text{orig}/I_\text{filtered}$, is more than a threshold of 1.3. Lastly, we mask all pixels within a $5 \times 5$ square centered on these identified pixels. We note that the parameters used here are determined for the specific sample in this study; for other samples, it may be necessary to adjust these parameters to optimize the results.

\begin{figure}
    \centering
    \includegraphics{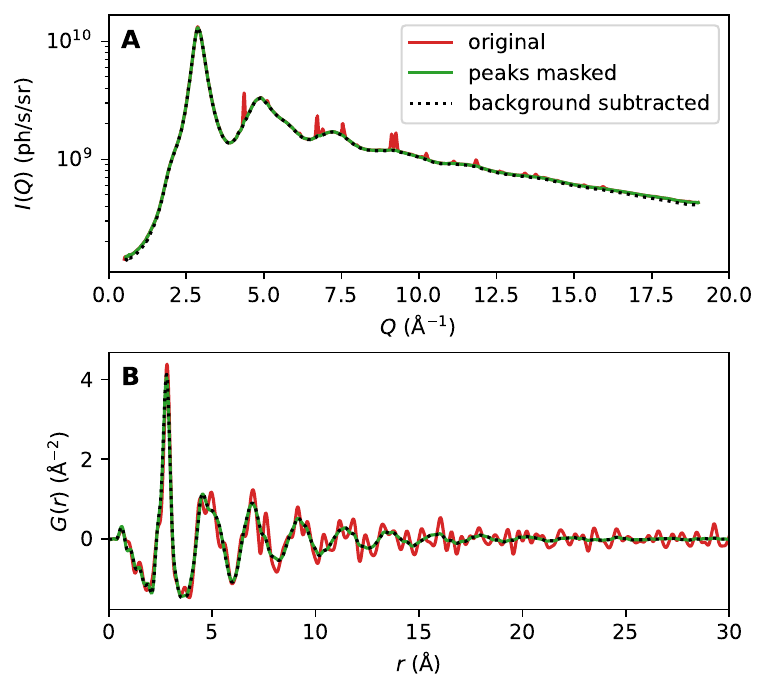}
    \caption{Effects of Bragg peak masking and background subtraction. (A) Azimuthally integrated intensity, $I(Q)$, of the original image (red), after peak masking (green), and after background subtraction (black dotted line). The original image is the same as in Figure~\ref{fig:background}. (B) Corresponding PDFs.}
    \label{fig:peak_masking}
\end{figure}

Figure~\ref{fig:peak_masking} shows an example of the result of peak masking. The original image is the same as the one shown as ``sample'' in Figure~\ref{fig:background} and contains obvious Bragg peaks. These Bragg peaks lead to significant oscillations in the calculated PDF that should not appear for an amorphous sample. After the peaks are masked following the procedure described above, both $I(Q)$ and $G(r)$ become smooth and typical of amorphous materials. We also demonstrate that the subtraction of the background shown in Figure~\ref{fig:background} does not lead to significant changes in $I(Q)$ or $G(r)$, as the results visibly overlap with the curves before this background subtraction in Figure~\ref{fig:peak_masking}.

We note that, for demonstration purposes, the example chosen here is not specifically optimized to reduce the intensity of the Bragg peaks. Nonetheless, the amount of masked pixels not adjacent to detector panel edges or beamstop shadow (both create sharp edges in the images which are identified by the algorithm) is on the order of \num{1.5e4}, less than 1.5\% of the valid pixels. Therefore, the peak masking does not cause significant degradation to the data quality.

\begin{acknowledgements}
We thank Denis Duran and Emmanuel Papillon for assistance with the setup at ID15A.
We thank Dr.\ Carlo Scian for technical support in the construction and testing of the vacuum chamber.
We thank Prof.\ Eloi Pineda for providing the sample used in this study.
\end{acknowledgements}

\begin{funding}
This project is financed by the European Union – NextGenerationEU and by the STARS@UNIPD program. We acknowledge the ESRF for provision of synchrotron radiation facilities under proposal number SC-5581.
D.L., L.P., F.D. and G.M. acknowledge support by the project GLAXES ERC-2021-ADG (Grant Agreement No. 101053167) funded by the European Union.
\end{funding}

\ConflictsOfInterest{The authors declare no conflicts of interest.
}

\DataAvailability{The data supporting the results reported in the present work can be found within the article. Raw data are obtained at beamline ID15A of the ESRF and can be found at https://doi.org/10.15151/ESRF-ES-2005462101. Further data can be provided by the corresponding author, P.S., upon reasonable request.}


\end{document}